\documentstyle[aps,preprint,tighten]{revtex}

\begin{document}

\preprint{\vbox{\hbox{TRI--PP--98--37}}}

\title{Comment on 'Induced Pseudoscalar Coupling Constant' by Il-Tong Cheon and
Myung Ki Cheoun (nucl-th/9811009)}

\author{Harold W. Fearing }
\address{TRIUMF, 4004 Wesbrook Mall, Vancouver, British Columbia, Canada 
         V6T 2A3}

\date{November 16, 1998}
\maketitle

\begin{abstract}
In a recent preprint Cheon and Cheoun have derived from a chiral model an
additional term, not usually appearing in the standard matrix element for
radiative muon capture. Using that term they generate a large correction to the
RMC spectrum which tends to resolve the problem caused by the too large value
of $g_P$ found in the TRIUMF RMC experiment. In this comment we observe first
that their extra term leads to an amplitude which is not gauge invariant and
second that such a term should be present, in a gauge invariant way, in an
earlier full chiral perturbation theory calculation, which however found
negligible differences from the standard approach.
\end{abstract}


A recent TRIUMF experiment \cite{TRIUMF} on radiative muon capture (RMC) on the
proton found a value of the induced pseudoscalar coupling constant $g_P$ which
was almost 1.5 times the value predicted by the Goldberger-Treiman
relation. This experiment was analyzed using the standard approach \cite{HWF}
and in particular the explicit formalism of Beder and Fearing \cite{Beder}. It
presents a puzzle since other measurements of $g_P$, particularly in non
radiative muon capture, seem to agree with the expected value.

Cheon and Cheoun \cite{Cheon} have proposed a possible solution to this. They
use a simple chiral model to generate an additional seagull type term which
they claim is not included in the standard approach. They then show that this
extra term has an appreciable effect on the photon spectrum which, at least
qualitatively, would solve the problem with $g_P$ presented by the TRIUMF data.

The purpose of this comment is to make some observations which suggest some
difficulties with this proposed solution to the problem.

The first observation is that the RMC amplitude generated by including this
extra term is not gauge invariant. The full amplitude is given by Eq. (25) of
Cheon and Cheoun \cite{Cheon}. It consists of the standard contributions $M_a,
... M_e$, which are identical to, for example, Eq. (1) of Ref. \cite{HWF}, plus
the new term $\Delta M_e$. It is well known that the standard terms are gauge
invariant, in fact the amplitude $M_e$ is obtained via a minimal substitution
on the other amplitudes and is included specifically to enforce gauge
invariance. Alternatively it is easy to check gauge invariance explicitly by
making the substitution $\epsilon \rightarrow k$ in the amplitudes $M_a,
... M_e$. The same substitution made in the new term $\Delta M_e$ leads to a
result proportional to $\overline{u}_n \gamma_5 k\!\!\!/ u_p$ which is not zero
in general, so that the full amplitude is not gauge invariant.

Gauge invariance is important because it forces the cancellation of similar
sized terms. One can see for example that gauge invariance of the $g_P$ terms
of the standard amplitude comes about via contributions from $M_a, M_b, M_d,$
and $M_e$ all of which are of the same order. Thus one might expect that a
fully gauge invariant amplitude should include additional terms similar to, and
of the same order as, $\Delta M_e$, with a priori unknown numerical
consequences.  But, in any case, one can not trust the numerical predictions
arising from a non gauge invariant amplitude.

The second observation is that this new term in principle would have been
included already in an earlier chiral perturbation theory (ChPT) calculation
\cite{Ando} which however found negligible effects relative to the standard
calculation.  ChPT starts with the most general Lagrangian satisfying chiral
symmetry. Since this new term was derived from a chiral model one would expect
that a term of the same structure would appear in the ChPT Lagrangian.  Ando
and Min \cite{Ando} recently performed a full, and presumedly gauge invariant,
ChPT calculation of RMC to $O(p^3)$. The low energy constants needed were all
determined either directly from experiment or via some sort of meson dominance
assumption. They found that the $O(p^3)$ terms were negligible and did not
solve the $g_P$ problem.

One can be a bit more explicit. In the notation of \cite{OMC,RMC} one can show
explicitly, without doing a full calculation, that in the usual ChPT
calculation there is a term even at leading order which would generate a
seagull amplitude with the structure of $\Delta M_e$. Furthermore there are
also the terms needed to make it gauge invariant. One can also look at the
heavy baryon ChPT reduction of the amplitude of the standard approach, simply
by reducing the spinors and gamma matrices as done in \cite{OMC,RMC} and see
that the term $\Delta M_e$ appears already in the standard amplitude, in $M_b$,
with the terms required for gauge invariance coming from the other diagrams.

The author would like to thank T.-S. Park for pointing out an error in the
original version of this comment and for some useful discussions, and Dong-Pil
Min also for some useful comments.  This work was supported in part by the
Natural Sciences and Engineering Research Council of Canada.

\end{document}